\documentclass[titlepage,12pt,letterpaper]{article}%
\usepackage{mathtools,amsthm,amssymb}
\usepackage{natbib}
\usepackage{hyperref}
\usepackage[center]{caption}
\usepackage{setspace}
\usepackage{epsfig}
\usepackage{graphics}
\usepackage{lscape}
\usepackage[english]{babel}
\usepackage{amsmath}
\usepackage{amsfonts}
\usepackage{amssymb}
\usepackage{graphicx}%
\providecommand{\U}[1]{\protect\rule{.1in}{.1in}}
\hypersetup{colorlinks=true,urlcolor=blue,citecolor=red,linkcolor=green}
\theoremstyle{plain}

\theoremstyle{plain}
\newtheorem* {corollary}{Corollary}
\theoremstyle{plain}

\theoremstyle{plain}

\theoremstyle{plain}

\theoremstyle{definition}

\theoremstyle{definition}

\theoremstyle{remark}

\theoremstyle{remark}

\theoremstyle{remark}

\begin{document}

\title{Rational Dialogues \thanks{We wish to thank Christina Pawlowitsch for giving
us the opportunity to write this paper.}}
\author{J. Geanakoplos \thanks{John Geanakoplos is a Professor of Economics at Yale
University and an External Professor at the Santa Fe Institute.
john.geanakoplos@yale.edu}
\and H. Polemarchakis
\thanks{\href{mailto:h.polemarchakis@warwick.ac.uk}{\texttt{h.polemarchakis@warwick.ac.uk}%
}}}
\date{May 5, 2023}
\maketitle

\begin{abstract}
Any finite conversation can be rationalized.

\bigskip

\bigskip


\bigskip


\end{abstract}

\bibpunct{(}{)}{;}{a}{,}{,}

In the summer of 1977 Herakles came to John very excited about a paper of Bob
Aumann's on common knowledge. We couldn't believe the paper, much less figure
it out. Our adviser Kenneth Arrow couldn't either. Even the title
\textquotedblleft\textit{Agreeing to Disagree}\textquotedblright\ seemed to
say the opposite of the paper's conclusion.

There is nothing more tantalizing than a paradox. As Aumann has managed with
other young students time and again, we were hooked. His teaching style builds
on paradoxes. He once described capitalism through a letter he had gotten from
his son about life in the Kibbutz. In the morning, the son had written, I do
something for my community and in the afternoon something for myself. Aumann
said, wouldn't it be better if by doing something for himself he was at the
same time doing something for his community? He described integration and the
fundamental theorem of calculus as showing that it is easier to solve many
hard problems than a single hard problem. At a conference in India, he was
asked by a reporter to say a word explaining Game Theory. Aumann replied that
the question reminded him of Nikita Khruschev's first press conference in
front of foreign journalists. A reporter asked Khrushchev to say a word about
the health of the Russian economy. Khrushchev said \textquotedblleft Good."
The reporter said he didn't literally mean one word, could Khruschev say two
words about the health of the Russian economy? Khrushchev replied
\textquotedblleft Not Good." Aumann continued by saying that in one word game
theory is about \textquotedblleft Interaction." In two words, it is about
\textquotedblleft Rational Interaction."

A parodox is something that sounds crazy, but looked at the right way makes
sense. For Aumann, paradoxes abound. In honor of Bob Aumann, we prove here
that paradoxes are ubiquitous. We show that any conversation, no matter how
crazy the opinions and the rejoinders sound, can be explained as the first
part of a dialogue between two perfectly rational interlocutors. Dialogues are
interactions. And they might all be rational interactions.

Turing famously suggested that one could distinguish a (nonrational) machine
from a man by engaging it in conversation and then letting a panel of judges
review the transcript and vote man or machine. As is becoming clearer today
with ChatGPT, and as our theorem suggests, it may not be as easy as Turing hoped.

Bob Aumann himself has often written that what is called irrational behavior
by behavioral economists might one day be better understood as rational
behavior in a complicated environment with constraints. Our theorem has a
similar flavor. Perhaps the most comforting aspect of our theorem is that it
provides some hope for our current troubled and polarized discourse.

\citet*{aumann76} defined \textit{common knowledge} and proved that consensus
is a necessary condition for common knowledge, that is, that people cannot
agree to disagree about the probability of an event. A \textit{Bayesian
dialogue} is a sequential exchange of beliefs about the probability of an
event. It is the prototype of a \textit{rational dialogue}. One of two
interlocutors states his belief, then the other responds with her belief,
perhaps informed or influenced by his stated belief. He then responds, perhaps
with a revision of his prior opinion (in view of her opinion), and then she
responds again, and so on. The dialogue is said to terminate at a time $T$ if
neither agent changes his or her mind thereafter. In
\citet*{geanakoplospolemarchakis82}, we proved that Bayesian dialogues must
always terminate, and that when they do, the agents are in
agreement.\footnote{We allowed for an arbitrary but finite state space.
\citet*{bacharach79} looked at Bayesian dialogues when information is normally
distributed. \citet*{nielsen84} considers dialogues with an uncountable number
of states.}

We show here that a third party, with access only to the transcript of a
dialogue, cannot be sure that any arbitrary finite sequence of alternating
opinions is not part of a Bayesian dialogue. If the transcript were infinitely
long, then it would necessarily terminate in agreement. We show that the
available finite transcript of opinions can always be continued to reach an
agreement in such a way that the whole dialogue from the beginning is rational.

Our argument covers the special case of a \textit{didactic dialogue}, in which
an expert is better informed than his interlocutor. The expert never changes
his opinion, but the interlocutor follows an arbitrary path. Some of Plato's
dialogues might be considered didactic dialogues in our sense. Socrates knows
the right answer to which he leads his interlocutor. Plato perhaps understood
our theorem in the sense that in some of his dialogues he has an interlocutor
of Socrates, such as Protagoras, appear at first to move further away from the
answer until eventually coming back to the right path.

Our theorem relies on one important premise. If an agent expresses absolute
certainty in her opinion, then her interlocutor must immediately agree.
Absolute certainty is tantamount to claiming a proof. If the interlocutor does
not agree, then one or the other cannot be rational. She can be 99.9999\%
certain of one thing, and then 99.9999\% certain of the opposite at the next
stage; as long as neither she nor he is 100\% certain, then whatever her
interlocutor and she say can be rationalized.

Loosely speaking, one can consider common knowledge and agreement as an
equilibrium, and the dialogue that leads to common knowledge as the adjustment
path. We are arguing that along the adjustment path, anything goes. This bears
an analogy with general competitive analysis. As follows from
\citet*{debreu74}, the Walrasian t\^{a}tonnement that leads to equilibrium, if
it does, is arbitrary.

\section*{The argument}

\subsection*{Bayesian Dialogues}

A \emph{Bayesian opinion framework} is defined by a finite probability space,
a subset, two partitions, and an agent,%
\[
(\Omega,\pi,A,P,Q,i),
\]
where $\Omega$ is a \emph{finite} set of states and $\pi$ is strictly positive
probability on $\Omega,$ and $A$ is a subset of $\Omega.$ The probability
$\pi$ is the common prior of two agents $p,q.$ $P$ and $Q$ are partitions of
$\Omega,$ corresponding to the two agents $p,q,$ defined by disjoint subsets
or cells $(P_{c})$ and $(Q_{d}),$ respectively. For any $\omega\in\Omega,$
$P(\omega)$ is defined as the unique cell $P_{c}$ containing, $\omega,$ and
likewise for $Q(\omega)${.} Finally, the agent $i\in\{p,q\}.$

The \emph{Bayesian opinion} of agent $i=p$ about the likelihood of $A,$
conditional on what $p$ knows, is defined by the function $i_{A}=p_{A}
:\Omega\rightarrow\lbrack0,1]$
\[
i_{A}(\omega)=p_{A}(\omega)=\frac{\pi(P(\omega)\cap A)}{\pi(P(\omega))},
\]
and, likewise, when $i=q,$
\[
i_{A}(\omega)=q_{A}(\omega)=\frac{\pi(Q(\omega)\cap A)}{\pi(Q(\omega))},
\]
defines $q$'s Bayesian opinion of the likelihood of $A$ conditional on what
$q$ knows.

If $i=p,$ then, after hearing $p$'s Bayesian opinion $p_{A},$ $q$ will revise
her understanding of the world, replacing $Q$ with $Q^{\prime}=Q\vee p_{A}
\ $defined by
\[
\lbrack Q\vee p_{A}](\omega)=Q(\omega)\cap\{\omega^{\prime}:p_{A}
(\omega^{\prime})=p_{A}(\omega)\}\text{ for all }\omega\in\Omega.
\]
Similarly, if $i=q,$ then after hearing $Q$'s Bayesian opinion $q_{A},$ $p$
will revise his understanding of the world, replacing $P$ with $P^{\prime
}=P\vee q_{A}\ $defined by
\[
\lbrack P\vee q_{A}](\omega)=P(\omega)\cap\{\omega^{\prime}:q_{A}%
(\omega^{\prime})=q_{A}(\omega)\}\text{ for all }\omega\in\Omega.
\]

Thus the Bayesian opinion framework $(\Omega,\pi,A,P,Q,p)$\ generates a unique
successor $(\Omega,\pi,A,P^{\prime},Q^{\prime},-p)=(\Omega,\pi,A,P,[Q\vee
p_{A}],q)$\ and the Bayesian opinion framework $(\Omega,\pi,A,P,Q,q)$%
\ generates a unique successor
\[
(\Omega,\pi,A,P^{\prime},Q^{\prime},-q)=(\Omega,\pi,A,[P\vee q_{A}],Q,p).
\]

It follows that any Bayesian opinion framework $(\Omega,\pi,A,P,Q,i)$
\ generates a uniquely defined infinite sequence of Bayesian opinion
frameworks
\[%
\begin{array}
[c]{c}%
(\Omega,\pi,A,P,Q,i)=(\Omega,\pi,A,P_{1},Q_{1},i_{1}),\\
\quad(\Omega,\pi,A,P_{2},Q_{2},i_{2}=-i),\\
(\Omega,\pi,A,P_{3},Q_{3},i_{3}=i),\\
\quad(\Omega,\pi,A,P_{4},Q_{4},i_{4}=-i),\\
\ldots
\end{array}
\]
in which, at each period, one agent $i$ gives his opinion based on his
partition at that time, and then in the next period the \emph{other} agent
$-i$\ gives her opinion based on her previous partition revised in light of
the previous opinion expressed by him. We call this whole infinite sequence a
\emph{Bayesian Dialogue} $(\Omega,\pi,A,P,Q,i)_{\infty}.$

\subsection*{Dialogues and Rational Dialogues}

Bayesian dialogues contain many counterfactual statements, covering opinions
conditional on all possible worlds $\omega\in\Omega.$ In reality we typically
only hear about a \emph{finite} number of \emph{actual} opinions. We define a
\emph{dialogue} as a finite sequence of opinions or beliefs $(b_{1}%
,b_{2},....,b_{T})$ with $b_{t}\in\lbrack0,1]$ for all $t.$ Once we specify a
fixed state of the world $\omega^{\ast}\in\Omega,$ every Bayesian dialogue
generates an infinite sequence of beliefs $(\Omega,\pi,A,P,Q,i,\omega^{\ast
})_{\infty}\equiv(r_{1},r_{2},...)$ where $r_{t}=i_{t,A}(\omega^{\ast})$ is
the opinion expressed by the opining agent at time $t$ for state $\omega
^{\ast}.$ We call this infinite sequence of opinions a Bayesian dialogue at a
fixed state. A \emph{rational dialogue} is any finite sequence of beliefs
$(r_{1},r_{2},...,r_{T})$ that can be realized as the first part of a Bayesian
dialogue at a fixed state.\bigskip

\citet*{geanakoplospolemarchakis82} showed that in any Bayesian dialogue at a
fixed state, $(r_{1},r_{2},...),$ there must be a finite time $T$ by which
consensus is reached, $r_{t}=r_{T}$ for all $t\geq T.$ Moreover, because
Bayesian rational agents believe in each other's rationality, if for some $T,$
$r_{T}\in\{0,1\}$, meaning one of the agents is absolutely certain and will
never change his/her mind, then consensus must have already been reached by
time $T.$

We say that the event $E\subset\Omega$ is common knowledge at $\omega^{\ast
}\in E$ if $P(\omega)\cup Q(\omega)\subset E$ for all $\omega\in E.$ It is
evident that if $E\subset\Omega$ is common knowledge at $\omega^{\ast},$\ then
the Bayesian dialogue at a fixed state $\omega^{\ast}$ $(\Omega,\pi
,A,P,Q,i_{A},\omega^{\ast})_{\infty}\equiv(r_{1},r_{2},...)$ does not depend
on any $P(\omega),Q(\omega)$ or $\pi(\omega)$ for $\omega\notin E.$

\subsection*{Irrational Dialogues?}

Are there dialogues $(b_{1},...,b_{T})$ that look so crazy that they could not
be the beginnings of a rational dialogue? Geanakoplos and Polemarchakis showed
that given any positive integer $n,$ there is a Bayesian dialogue
$(c,d,c,d,...,$ $c,d,c,c,...)$ in which one agent obstinately maintains the
opinion $c$ while the other maintains $d\neq c,$ and then, suddenly, after $n$
such alternations, consensus is reached at $c.$

In an unpublished paper \citet*{polemarchakis16} showed that any dialogue
could be rational. \citet*{lehreretal22} extended the theorem to infinite
dialogues. The following theorem gives a similar result to
\citet*{polemarchakis16} but in a slightly different setting and with a
different proof.

There is one property that must hold for any rational dialogue, because
rationality presumes both agents are rational and know that both are rational.
If one of the agents is certain, then the other must immediately agree.
Certainty is tantamount to claiming a proof, and if the other does not agree,
one of the two interlocutors must not be rational. \bigskip

\textbf{Definition}: The dialogue $(b_{1},...,b_{T})$ violates \emph{certainty
acquiescence} if for some $t<T,$ $b_{t}\in\{0,1\},$\emph{ }yet $b_{t+1}\neq
b_{t}.$\emph{ }\bigskip

Needless to say, if in the dialogue $(b_{1},...,b_{T})$ nobody expresses
absolute certainty, then the dialogue does not violate certainty acquiescence.
The opinions could bounce around arbitrarily, as long as none hit $0$ or $1$.

\paragraph{Theorem}

\emph{Let }%
\[
(b_{1},...,b_{T})
\]
\emph{be an arbitrary dialogue that does not violate certainty acquiescence.
Then }$(b_{1},...,b_{T})$\emph{ is a rational dialogue generated by some
}$(\Omega,\pi,A,P,Q,p_{A},\omega^{\ast})_{\infty}.$\emph{ Moreover, consensus
is reached at time }$T$\emph{ at }$b_{T}.$

\paragraph{Proof}

The proof is by backward induction. Suppose $T=1.$ Suppose $0<b_{T}<1.$ Let
$\Omega=\{y,n\},$ and let $A=\{y\}.$ Let $\pi(y)=b_{T},$ and $\pi(n)=1-b_{T}.$
Let $P=Q=\{\{y,n\}\}.$ Let $\omega^{\ast}=y.$ Clearly consensus is reached at
$b_{T}$ because both agents have the same information.

If $b_{T}=1$, delete the point $n,$ and continue as above. If $b_{T}=0$, let
$\Omega=\{y,n\},$ and let $A=\{y\}$ and let $P=Q=\{\{y\},\{\omega^{\ast
}=n\}\}$ and let $\pi(y)=\pi(n)=1/2.$ Clearly in all three cases the rational
dialogue reaches consensus on the first step at $b_{T},$ no matter which agent
is the first opiner.

Now suppose the theorem has been proved for all $T\leq n$ and let
$(b_{1},b_{2},...,$ $b_{T})$ be given for $T=n+1.$ By the induction hypothesis
we can find a Bayesian dialogue at a fixed state $(\Omega,\pi,A,P,Q,q,\omega
^{\ast})_{\infty}=(b_{2},...,b_{T},...)$, with consensus at $b_{T},$ in which
$q$ is the first speaker (with opinion $b_{2}).$ We shall now define
\[
(\Omega^{\ast},\pi^{\ast},A^{\ast},P^{\ast},Q^{\ast},p,\omega^{\ast\ast
})_{\infty}\equiv(r_{1},...,r_{T},...)
\]
with consensus at time $T$ at $r_{T},$ with $r_{t}=b_{t}$ for $t=1,...,T,$ in
which $p$ is the first speaker. If $b_{1}=1,$ then by certainty acquiescence,
$b_{t}=1$ for all $t$ and we can rationalize that with the Bayesian dialogue
in the second paragraph, and similarly if $b_{1}=0$.

So suppose $0<b_{1}<1.$ Define $\Omega^{\ast}$ and $P^{\ast}$ by adding to
$\Omega$\ two extra points $y_{c},n_{c}$ for \emph{each} partition cell
$P_{c},$ so $P_{c}^{\ast}=P_{c}\cup\{y_{c},n_{c}\}.$ The partition $Q^{\ast}$
adds two cells to those already in $Q,$ namely $Q_{y}^{\ast}$ consisting of
all the $y_{c}$, and the other $Q_{n}^{\ast}$ consisting of all the $n_{c}.$
$A^{\ast}$ extends $A$ by including also all the $y_{c}.$ This situation is
depicted in Diagrams 1 and 2.%

\begin{center}
\includegraphics{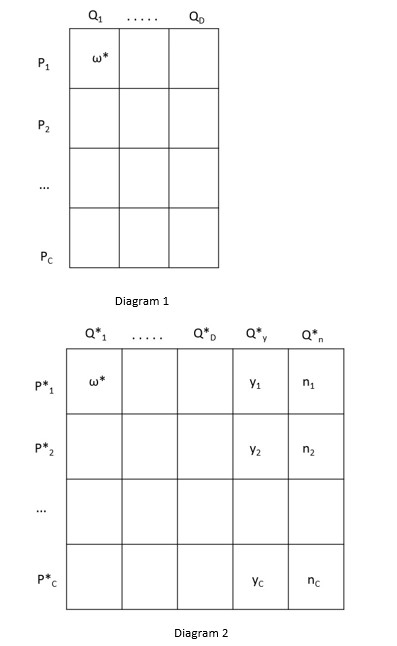}
\end{center}

The crucial step is to note that for \emph{every} partition cell $P_{c},$
there exists numbers $0<\pi(y_{c})<1$ and $0<\pi(n_{c})<1$ such that
\[
b_{1}=\frac{\pi(A\cap P_{c})+\pi(y_{c})}{\pi(P_{c})+\pi(y_{c})+\pi(n_{c})}%
\]
This extends the probability measure $\pi$ to a measure on all of
$\Omega^{\ast}.$\ Define $\pi^{\ast}$ by rescaling the $\pi$ (over all
$\Omega^{\ast})$ so that they add to $1.$ Observe that the rescaling in
numerator and denominator cancel, so for \emph{all} $P_{c},$%
\[
b_{1}=\frac{\pi(A\cap P_{c})+\pi(y_{c})}{\pi(P_{c})+\pi(y_{c})+\pi(n_{c}%
)}=\frac{\pi^{\ast}(A^{\ast}\cap P_{c}^{\ast})}{\pi^{\ast}(P_{c}^{\ast})}.
\]
Take $\omega^{\ast\ast}=\omega^{\ast}.$ This completes the definition of the
Bayesianl dialogue announced above.

At the first step agent $p$ announces
\[
p_{A}(\omega^{\ast\ast})\equiv r_{1}=\frac{\pi^{\ast}(A^{\ast}\cap P^{\ast
}(\omega^{\ast\ast}))}{\pi^{\ast}(P^{\ast}(\omega^{\ast\ast}))}=b_{1}%
\]
and reveals nothing, because, as noted, for every partition cell $p$ would
announce the same, Hence at the next step $Q^{\ast\prime}=Q^{\ast}.$ But
$\omega^{\ast\ast}=\omega^{\ast}\in\Omega,$ hence by construction $Q^{\ast
}(\omega^{\ast\ast})=Q(\omega^{\ast\ast})=Q(\omega^{\ast})\subset\Omega.$ Thus
$q$ then announces
\[
\frac{\pi^{\ast}(A^{\ast}\cap Q^{\ast\prime}(\omega^{\ast\ast}))}{\pi^{\ast
}(Q^{\ast\prime}(\omega^{\ast\ast}))}=\frac{\pi(A\cap Q(\omega^{\ast}))}%
{\pi(Q(\omega^{\ast}))}=b_{2},
\]
where the last equality follows from the induction hypothesis and the fact
that $\pi^{\ast}$\ scales $\pi$. If $b_{2}\in\{0,1\},$ then this rational
dialogue, like all rational dialogues, repeats $b_{2}$ thereafter, reproducing
the given dialogue which, by certainty acquiescence would have repeat $b_{2}$ thereafter.

So suppose $0<b_{2}<1.$ Then this announcement of $b_{2}$ makes it common
knowledge that $\omega^{\ast\ast}\in\Omega,$ because had $q$ seen partition
cell $Q_{y}^{\ast}$ or $Q_{n}^{\ast},$ she would have announced $1$ or $0$
instead of $b_{2}.$ Thus the Bayesian dialogue at a fixed state $(\Omega
^{\ast},\pi^{\ast},A^{\ast},P^{\ast},Q^{\ast},p,\omega^{\ast\ast})$ begins
with $b_{1}$ and then from step 2 onwards proceeds as the Bayesian dialogue
with a fixed state $(\Omega,\pi,A,P,Q,q,\omega^{\ast})$.$\blacksquare$

\section*{Remarks}

\paragraph{An Example of the Construction}

The constructive argument above can generate any dialogue, no matter how
curious. For example, the two agents could agree with each other on say the
probability $1/4$ for many iterations, and then suddenly jump to consensus at
$3/4$.

We give the construction for the dialogue%
\[
(\frac{1}{4},\frac{1}{4},\frac{1}{4},\frac{1}{4},\frac{3}{4},\frac{3}{4})
\]
in the matrix below, where each $y$ corresponds to a different state in $A$
and each $n$ corresponds to a different state in $\Omega\backslash A,$ and the
numbers in the brackets are measures for the corresponding states. The
probability measure that is the common prior of the agents is given by
normalizing these measures to add to one, namely the numbers in brackets
divided by $13\frac{2}{3}.$ Observe that conditional probabilities are not
affected by replacing a measure with any scalar multiple of the measure. The
partition of agent $p$ consists of the rows of the matrix, and the partition
of agent $q$ corresponds to the columns of the matrix. The state of nature
$\omega^{\ast}$ is the $y$ in the top left corner. Notice that the top left
cell of the matrix is the only one containing two points.%
\[%
\begin{array}
[c]{ccccc}%
y[\frac{3}{4}],n[\frac{1}{4}] & y[0] & n[2] & y[0] & n[0]\\
y[0] & y[\frac{1}{4}] & n(\frac{3}{4}) & y[0] & n[0]\\
n[2] & y[\frac{2}{3}] & n[0] & y[0] & n[0]\\
y[0] & y[0] & y[1] & y[0] & n[3]\\
n[0] & n[\frac{11}{4}] & n[\frac{1}{4}] & y[1] & n[0]
\end{array}
\]

The reader can check that $p$ will announce $1/4=(3/4)/(3/4+1/4+2),$ revealing
nothing since the conditional probability of $A$ given any row is exactly
$1/4$. Then $q$ will announce $1/4$, since the conditional probability of $A$
in the left most column is $1/4$. That reveals precisely that $\ \omega^{\ast
}$ is not in one of the last two columns, since they would have led to the
announcements of $1$ or $0$. With this information, $p$ still says $1/4$,
since that is the conditional probability of $A$ given the top row without its
last two elements. This announcement reveals that $\omega^{\ast}$ is not in
the bottom two rows, since they would have led to the announcements of $1$ or
$0$. Agent $q$ responds to this by still saying $1/4$ since that is the
probability of $A$ given the first column without its last two elements. That
reveals to $p$ that $\omega^{\ast}$ is not the second or third columns, since
they would have led to the announcements of $1$ or $0$. With this information,
$p$ finally says $3/4$. This reveals that $\omega^{\ast}$ is in the top left
cell, and gets agreement from $q$ at $3/4$.

The measure makes clear how the probabilities were constructed by backward
induction. The top left cell is first in the construction. If that cell were
common knowledge, $p$ and $q$ would agree on $3/4$, giving the last two
opinions $(3/4,3/4)$ in the dialogue. Next we add the second and third
elements of the first column. The measures assigned to $y$ and $n$ induce $q$
to assign conditional probability of $1/4$ to seeing this part of the first
column. Thus we can generate the dialogue $(1/4,3/4,3/4).$%
\[%
\begin{array}
[c]{c}%
y[\frac{3}{4}],n[\frac{1}{4}]
\end{array}
\rightarrow%
\begin{array}
[c]{c}%
y[\frac{3}{4}],n[\frac{1}{4}]\\
y[0]\\
n[2]
\end{array}
\rightarrow
\]

Next we added the second and third columns of the first three rows, assigning
the measures to make sure that player $p$ gives conditional probability of $A$
of $1/4$ to each row so far constructed.
\[%
\begin{array}
[c]{ccc}%
y[\frac{3}{4}],n[\frac{1}{4}] & y[0] & n[2]\\
y[0] & y[\frac{1}{4}] & n(\frac{3}{4})\\
n[2] & y[\frac{2}{3}] & n[0]
\end{array}
\rightarrow
\]
This gives us a dialogue $(1/4,1/4,3/4,3/4).$Next we move to add the fourth
and fifth rows, as indicated below, so that $q$ assigns the same probability
$1/4$ to $A$ in each column. This gives us a dialogue $(1/4,1/4,1/4,3/4,3/4).$
Finally we add the last two columns so that $p$ gives conditional probability
of $A$ of $1/4$ to each row, giving us the whole dialogue
$(1/4,1/4,1/4,1/4,3/4,3/4)$.
\[%
\begin{array}
[c]{ccc}%
y[\frac{3}{4}],n[\frac{1}{4}] & y[0] & n[2]\\
y[0] & y[\frac{1}{4}] & n(\frac{3}{4})\\
n[2] & y[\frac{2}{3}] & n[0]\\
y[0] & y[0] & y[1]\\
n[0] & n[\frac{11}{4}] & n[\frac{1}{4}]
\end{array}
\rightarrow%
\begin{array}
[c]{ccccc}%
y[\frac{3}{4}],n[\frac{1}{4}] & y[0] & n[2] & y[0] & n[0]\\
y[0] & y[\frac{1}{4}] & n(\frac{3}{4}) & y[0] & n[0]\\
n[2] & y[\frac{2}{3}] & n[0] & y[0] & n[0]\\
y[0] & y[0] & y[1] & y[0] & n[3]\\
n[0] & n[\frac{11}{4}] & n[\frac{1}{4}] & y[1] & n[0]
\end{array}
\]

\paragraph{Experts}

We described a didactic dialogue as one in which an expert leads a student
through a conversation. At a state of the world, $\omega^{\ast},$ individual 1
is an expert concerning the event $A$ if no information in the join (coarsest
refinement of the partitions of the individuals) would cause him to alter his beliefs.

A \textit{didactic dialogue }is a dialogue
\[
(\bar{q},q^{2},\ldots,\bar{q},q^{2t},\ldots,\bar{q}),\quad0\leq\bar
{q},...,q^{T}\leq1,
\]
where the opinion at $t$ odd is unchanging, $q^{2t+1}=\bar{q},$ and at $t$
even, $q^{2t},$ is arbitrary but never 1 or 0.

\begin{corollary}
Any didactic dialogue $(\bar{q},q^{2},\ldots,\bar{q},q^{2t},\ldots,\bar{q})$
is a rational dialogue.
\end{corollary}

The following matrix of states and measures displays a Bayesian opinion
framework, with the row player opining first. At the fixed state given by $y$
in the to left, this gives a Bayesian dialogue with the opinions%
\[
(\frac{3}{4},\frac{1}{4},\frac{3}{4},\frac{1}{4},\frac{3}{4},\frac{3}%
{4},....)
\]%
\[%
\begin{array}
[c]{ccccc}%
y[\frac{3}{4}],n[\frac{1}{4}] & y[0] & n[0] & y[0] & n[0]\\
y[0] & y[\frac{3}{4}] & n(\frac{1}{4}) & y[0] & n[0]\\
n[2] & y[6] & n[0] & y[0] & n[0]\\
y[0] & y[0] & y[\frac{1}{12}] & y[0] & n[\frac{1}{36}]\\
n[0] & n[\frac{81}{4}] & n[0] & y[\frac{243}{4}] & n[0]
\end{array}
\]

\paragraph{Silence}

Here, a dialogue is an alternating sequence of opinions. Formally, an
interlocutor cannot remain silent when it is her turn to speak. One can
interpret silence by an interlocutor at $t$ as the repetition of her opinion
at $t-2:$ that is, $b_{t}=b_{t-2}.$ Thus if our tape contains only the opinion
of one agent that is changing over time, we can interpret it as a conversation
with an expert who constantly repeats the same opinion.

\bibliographystyle{plainnat}
\bibliography{rational-bibliography}

\begin{thebibliography}{7}
\providecommand{\natexlab}[1]{#1}
\providecommand{\url}[1]{\texttt{#1}}
\expandafter\ifx\csname urlstyle\endcsname\relax
  \providecommand{\doi}[1]{doi: #1}\else
  \providecommand{\doi}{doi: \begingroup \urlstyle{rm}\Url}\fi

\bibitem[Aumann(1976)]{aumann76}
R.~J. Aumann.
\newblock Agreeing to disagree.
\newblock \emph{Annals of Statistics}, 4:\penalty0 1236--1239, 1976.

\bibitem[Bacharach(1979)]{bacharach79}
M.~Bacharach.
\newblock Normal bayesian dialogues.
\newblock \emph{Journal of the American Statistical Association}, 74:\penalty0
  837--846, 1979.

\bibitem[Debreu(1974)]{debreu74}
G.~Debreu.
\newblock Excess demand functions.
\newblock \emph{Journal of Mathematical Economics}, 1:\penalty0 000--000, 1974.

\bibitem[{Di Tillio} et~al.(2022){Di Tillio}, Lehrer, and Samet]{lehreretal22}
A.~{Di Tillio}, E.~Lehrer, and D.~Samet.
\newblock Monologues, dialogues and commom priors.
\newblock \emph{Theoretical Economics}, 17:\penalty0 587--615, 2022.

\bibitem[Geanakoplos and Polemarchakis(1982)]{geanakoplospolemarchakis82}
J.~D. Geanakoplos and H.~Polemarchakis.
\newblock We cannot disagree forever.
\newblock \emph{Journal of Economic Theory}, 28:\penalty0 192--200, 1982.
\newblock URL \url{http://www.polemarchakis.org/a16-cdf.pdf}.

\bibitem[Nielsen(1984)]{nielsen84}
L.~T. Nielsen.
\newblock Common knowledge, communication, and convergence of beliefs.
\newblock \emph{Mathematical Social Sciences}, 8:\penalty0 1--14, 1984.

\bibitem[Polemarchakis(2016)]{polemarchakis16}
H.~Polemarchakis.
\newblock Rational dialogs.
\newblock Unpublished manuscript, 2016.
\newblock URL \url{http://www.polemarchakis.org/u03-bad.pdf}.

\end{thebibliography}

\end{document}